\documentclass[aps,pra,twocolumn,superscriptaddress]{revtex4-1}

\usepackage{graphicx}
\usepackage{amsmath}

\begin{document}

\title{Second-order temporal interference of two independent light beams at an asymmetrical beam splitter}

\author{Jianbin Liu}
\email[]{liujianbin@xjtu.edu.cn}
\affiliation{Electronic Materials Research Laboratory, Key Laboratory of the Ministry of Education \& International Center for Dielectric Research, Xi'an Jiaotong University, Xi'an 710049, China}

\author{Jingjing Wang}
\affiliation{Electronic Materials Research Laboratory, Key Laboratory of the Ministry of Education \& International Center for Dielectric Research, Xi'an Jiaotong University, Xi'an 710049, China}

\author{Zhuo Xu}
\affiliation{Electronic Materials Research Laboratory, Key Laboratory of the Ministry of Education \& International Center for Dielectric Research, Xi'an Jiaotong University, Xi'an 710049, China}

\date{\today}

\begin{abstract}
The second-order temporal interference of classical and nonclassical light at an asymmetrical beam splitter is discussed based on two-photon interference in Feynman's path integral theory. The visibility of the second-order interference pattern is determined by the properties of the superposed light beams, the ratio between the intensities of these two light beams, and the reflectivity of the asymmetrical beam splitter. Some requirements about the asymmetrical beam splitter have to be satisfied in order to ensure that the visibility of the second-order interference pattern of nonclassical light beams exceeds classical limit. The visibility of the second-order interference pattern of photons emitted by two independent single-photon sources is independent of the ratio between the intensities. These conclusions are important for the researches and applications in quantum optics and quantum information when asymmetrical beam splitter is employed.
\end{abstract}

\maketitle

\section{Introducntion}\label{introduction}

Beam splitter (BS) is a simple yet important element in classical optics \cite{born}, quantum optics \cite{mandel-book}, and quantum information \cite{chuang-book}. Symmetrical BS was assumed in most of the existed studies in order to simplify the calculations \cite{mandel-book,chuang-book}. However, asymmetrical BS is more general than symmetrical BS, since it is difficult to produce symmetrical BS in practice. Further more, asymmetrical BS has important applications in quantum cryptography \cite{bennett-1992}, multiphoton de Broglie wavelength measurement \cite{wang-2005,liu-2007,liu-2008,resch-2007}, filtering out photonic Fock states \cite{sanaka-2006}, and other interesting applications \cite{wang-2004, fan-2006,jacques-2008,ou-2008,wittmann-2008,liang-2010,li-2012}. It will be helpful to understand how different the interference patterns are for symmetrical and asymmetrical beam splitters. There were studies about the properties of both symmetrical and asymmetrical beam splitters in quantum theory \cite{prasda-1987,ou-1987,frarn-1987}. However, systematical study about the second-order interference of two independent light beams at an asymmetrical BS is still missing. In this paper, we will employ two-photon interference theory to study this topic and show how the visibility of the second-order interference pattern is influenced by the superposed light beams and asymmetrical BS.

Although both classical and quantum theories can be employed to calculate the second-order interference of classical light, only quantum theory is valid when nonclassical light is employed \cite{glauber-1963,glauber-1963-1,sudarshan-1963}.  We have employed two-photon interference theory to discuss the second-order interference of light at a symmetrical BS \cite{liu-2010,liu-2013,liu-EPL,liu-submitted,liu-2015}, which is helpful to understand the physics behind the mathematical calculations. The same method will be employed to calculate the second-order interference of two independent light beams at an asymmetrical BS.

The following parts are organized as follows. In Sect. \ref{theory}, we will calculate the second-order interference of different types of light beams superposed at an asymmetrical BS based on the two-photon interference in Feynman's path integral theory. The discussions and conclusions are in Sects. \ref{discussion} and \ref{conclusion}, respectively.

\section{Theory}\label{theory}

\begin{figure}[htb]
    \centering
    \resizebox{0.3 \textwidth}{!}
   {\includegraphics{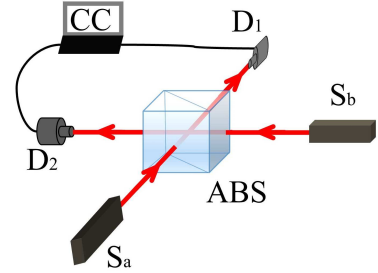}}
    \caption{The scheme for the second-order interference of two independent light beams at an asymmetrical BS. S$_a$ and S$_b$ are two independent light sources. ABS is an asymmetrical beam splitter. D$_1$ and D$_2$ are two single-photon detectors. CC is two-photon coincidence counting detection system.
    }\label{setup}
\end{figure}

The scheme for the second-order interference of two independent light beams at an asymmetrical BS is shown in Fig. \ref{setup}. S$_a$ and S$_b$ are two independent light sources, which can emit classical or nonclassical light. ABS is an asymmetrical beam splitter. D$_1$ and D$_2$ are two single-photon detectors. CC is two-photon coincidence counting detection system. The optical distances via ABS between S$_a$ and D$_1$, S$_a$ and D$_2$, S$_b$ and D$_1$, S$_b$ and D$_2$ are all assumed to be equal.

There are three different ways to trigger a two-photon coincidence count in Fig. \ref{setup}. One is these two photons are both emitted by S$_a$. The second way is these two photons are both emitted by S$_b$. The third way is these two photons are emitted by S$_a$ and S$_b$, respectively. The intensities of the light beams emitted by S$_a$ and S$_b$ are $I_a$ and $I_b$, respectively. The reflectivity and transmittivity of ABS are $R$ and $T$, respectively. The sum of $R$ and $T$ is 1 for a lossless BS \cite{prasda-1987,ou-1987,frarn-1987}. The probability for the photon detected by D$_1$ coming from S$_a$ is
\begin{equation}\label{p1a}
P_{1a}=\frac{I_a T}{I_a T+I_b R}.
\end{equation}
$P_{1b}$, the probability for the photon detected by D$_1$ coming from S$_b$, is $1-P_{1a}$. The probability for the photon detected by D$_2$ coming from S$_a$ is
\begin{equation}\label{p1b}
P_{2a}=\frac{I_a R}{I_a R+I_b T}.
\end{equation}
$P_{2b}$, the probability for the photon detected by D$_2$ coming from S$_b$, is $1-P_{2a}$. With the preparation above, we can calculate the probability for the three different ways to trigger a two-photon coincidence count in Fig. \ref{setup}. The probability for these two photons coming from S$_a$ is $P_{1a}P_{2a}$. The probability for these two photons coming from S$_b$ is $P_{1b}P_{2b}$. The probability for these two photons coming from S$_a$ and S$_b$, respectively, is $P_{1a}P_{2b}+P_{1b}P_{2a}$. The sum of these three probabilities equals 1.

For simplicity, we assume S$_a$ and S$_b$ are two point sources. The Feynman's photon propagator for point light source is  \cite{peskin}
\begin{eqnarray}\label{propagator}
K_{\alpha \beta}=\frac{\exp [-i(\vec{k}_{\alpha\beta}\cdot
\vec{r}_{\alpha\beta}-2\pi \nu_{\alpha}
t_{\beta})]}{r_{\alpha\beta}},
\end{eqnarray}
which is the same as Green function for a point light source in classical optics \cite{born}. $\vec{k}_{\alpha\beta}$ and $\vec{r}_{\alpha\beta}$ are the wave and position vectors of the photon emitted by S$_\alpha$ and detected at D$_\beta$, respectively. $r_{\alpha\beta}=|\vec{r}_{\alpha\beta}|$ is the
distance between S$_\alpha$ and D$_\beta$. $\nu_{\alpha}$ and $t_{\beta}$ are the frequency and time for the photon that is emitted by S$_\alpha$ and detected at D$_\beta$, respectively ($\alpha=a$ and $b$, $\beta=1$ and 2).

\subsection{The second-order interference of laser and thermal light}\label{lt-subsection}

We will first calculate the second-order interference of laser and thermal light in Fig. \ref{setup}. S$_a$ is assumed to be a thermal light source and S$_b$ is assumed to be a single-mode laser light source. If the photons emitted by these two sources are indistinguishable, these three different ways to trigger a two-photon coincidence count are indistinguishable. Based on the superposition principle in Feynman's path integral theory \cite{feynman-p}, the second-order coherence function in Fig. \ref{setup} is \cite{liu-2013,liu-EPL,liu-submitted}
\begin{eqnarray}\label{g2-lt}
&&G^{(2)}_{lt}(\vec{r}_1,t_1;\vec{r}_2,t_2)\nonumber\\
&&=\langle|\sqrt{P_{1a}P_{2a}}(\frac{1}{\sqrt{2}}e^{i\varphi_{a}}K_{a1}e^{i(\varphi_{a}'+\frac{\pi}{2})}K_{a2}\nonumber\\
&&+\frac{1}{\sqrt{2}}e^{i(\varphi_{a}+\frac{\pi}{2})}K_{a2}e^{i\varphi_{a}'}K_{a1})\nonumber\\
&&+\sqrt{P_{1b}P_{2b}}e^{i(\varphi_{b}+\frac{\pi}{2})}K_{b1}e^{i\varphi_{b}'}K_{b2}\nonumber\\
&&+\sqrt{P_{1a}P_{2b}}e^{i\varphi_{a}''}K_{a1}e^{i\varphi_{b}''}K_{b2}\nonumber\\
&&+\sqrt{P_{1b}P_{2a}}e^{i(\varphi_{b}''+\frac{\pi}{2})}K_{b1}e^{i(\varphi_{a}''+\frac{\pi}{2})}K_{a2}|^2\rangle.
\end{eqnarray}
Where $\langle...\rangle$ is ensemble average by taking all the possible phases into consideration. $\varphi_\alpha$ is the phase of photon emitted by S$_\alpha$ ($\alpha=a$ and b). The extra phase $\pi/2$ is due to the photon reflected by a BS will gain an extra phase comparing to the transmitted one. The first two terms on the righthand side of Eq. (\ref{g2-lt}) correspond to two detected photons are both emitted by the thermal light source, S$_a$. There are two indistinguishable alternatives for two photons in thermal light to trigger a two-photon coincidence count \cite{valencia-2005}. The third term on the righthand side of Eq. (\ref{g2-lt}) corresponds to two detected photons are both emitted by the laser light source, S$_b$. There is only one alternative \cite{liu-2013,liu-EPL,liu-submitted}. The last two terms on the righthand side of Eq. (\ref{g2-lt}) correspond to two detected photons are emitted by S$_a$ and S$_b$, respectively.

\begin{figure}[htb]
    \centering
    \resizebox{0.4 \textwidth}{!}
    {\includegraphics{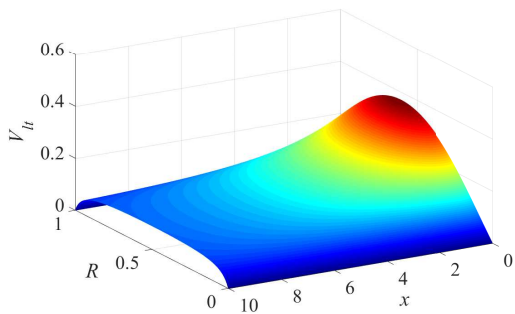}}
    \caption{Visibility of the second-order interference pattern of laser and thermal light. $V_{lt}$: visibility. $x$: ratio between $I_a$ and $I_b$. $R$: reflectivity.
    }\label{lt}
\end{figure}

The phases of photons in thermal light are random and the phases of photons in single-mode laser light are identical within the coherence time  \cite{loudon-book}. Taking these phase relations into consideration, Eq. (\ref{g2-lt}) can be simplified as
\begin{eqnarray}\label{g2-lt-1}
&&G^{(2)}_{lt}(\vec{r}_1,t_1;\vec{r}_2,t_2)\nonumber\\
&&=\frac{P_{1a}P_{2a}}{2}\langle|K_{a1}K_{a2}+K_{a2}K_{a1}|^2\rangle\nonumber\\
&&+P_{1b}P_{2b}\langle |K_{b1}K_{b2}|^2 \rangle \nonumber\\
&&+\langle |\sqrt{P_{1a}P_{2b}}K_{a1}K_{b2}-\sqrt{P_{1b}P_{2a}}K_{b1}K_{a2}|^2\rangle.
\end{eqnarray}
Substituting Eq. (\ref{propagator}) into Eq. (\ref{g2-lt-1}) and with the same method as the one in Refs. \cite{liu-2013,liu-EPL,liu-submitted}, it is straightforward to have one-dimension second-order temporal coherence function as
\begin{eqnarray}\label{g2-lt-2}
&&G^{(2)}_{lt}(t_1-t_2)\\
&\propto& 1+P_{1a}P_{2a}-2\sqrt{P_{1a}P_{1b}P_{2a}P_{2b}}\cos[2\pi \Delta \nu(t_1-t_2)],\nonumber
\end{eqnarray}
where quasi-monochromatic assumption is employed \cite{mandel-book}. These two detectors are assumed to be at symmetrical positions in order to concentrate on the temporal interference pattern. $\Delta \nu$ is the difference between the mean frequencies of the light beams emitted by S$_a$ and S$_b$, which equals $|\nu_a-\nu_b|$. The definition for the visibility of the second-order interference pattern is
\begin{equation}\label{vis-lt}
V=\frac{G^{(2)}_{max}-G^{(2)}_{min}}{G^{(2)}_{max}+G^{(2)}_{min}},
\end{equation}
where $G^{(2)}_{max}$ and $G^{(2)}_{min}$ are the maximal and minimal values of the second-order coherence function, respectively. The visibility of the second-order interference pattern of laser and thermal light in Fig. \ref{setup} is
\begin{equation}\label{vis-lt}
V_{lt}=\frac{2\sqrt{P_{1a}P_{1b}P_{2a}P_{2b}}}{1+P_{1a}P_{2a}}.
\end{equation}
Substituting Eqs. (\ref{p1a}) and (\ref{p1b}) into Eq. (\ref{vis-lt}), the visibility equals
\begin{equation}\label{vis-lt-1}
V_{lt}=\frac{2xR(1-R)}{(x+R-xR)(1-R+xR)+x^2R(1-R)},
\end{equation}
where $x$ equals $I_a/I_b$. The visibility of the second-order interference pattern of laser and thermal light depends on the ratio between the intensities of these two light beams and the reflectivity of the asymmetrical BS. Figure \ref{lt} shows how the visibility changes when the ratio ($x$) and reflectivity ($R$) vary. When $x$ equals $\sqrt{2}/2$ and $R$ equals $0.5$, the visibility gets its maximal value, $1/(\sqrt{2}+1)$, which is less than the limit of the visibility of the second-order interference pattern of classical light beams in Fig. \ref{setup} \cite{mandel-1983}.

\begin{figure}[htb]
    \centering
    \resizebox{0.4 \textwidth}{!}
    {\includegraphics{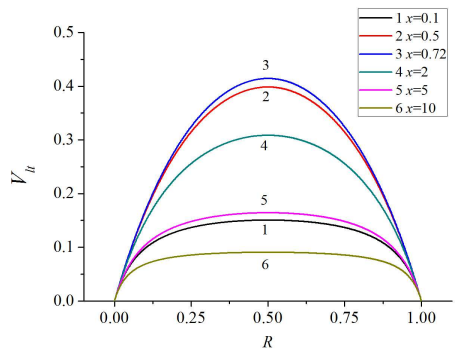}}
    \caption{Visibility vs. reflectivity for different ratios. The lines of 1-6 correspond to $x=$0.1, 0.5, 0.71, 2, 5, and 10, respectively.
    }\label{lt-x}
\end{figure}

The visibility in Fig. \ref{lt} is symmetrical about $R=0.5$, which can be clearly seen in Fig. \ref{lt-x}. The lines of 1-6 correspond to $x=$0.1, 0.5, 0.71, 2, 5, and 10, respectively. For a fixed $x$, the visibility gets its maximum when $R$ equals 0.5 and decreases when $R$ deviates from 0.5.

\begin{figure}[htb]
    \centering
    \resizebox{0.4 \textwidth}{!}
    {\includegraphics{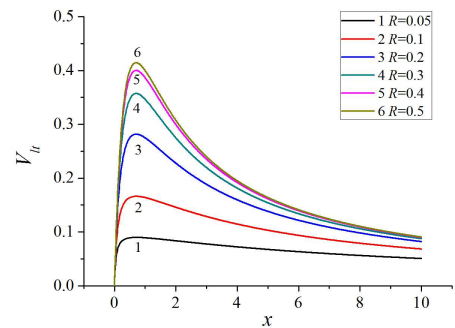}}
    \caption{Visibility vs. ratio for different reflectivities. The lines of 1-6 correspond to $x=$0.05, 0.1, 0.2, 0.3, 0.4, and 0.5, respectively.
    }\label{lt-R}
\end{figure}

Figure \ref{lt-R} shows how the visibility varies as $x$ changes for different reflectivity.  The visibility increases as $x$ increases from 0 to $\sqrt{2}/2$ and gets its maximum when $x$ equals $\sqrt{2}/2$. Then the visibility decreases as $x$ increases from $\sqrt{2}/2$. We only draw the situations when $x$ is not greater than 10 in Fig. \ref{lt-R}. It is easy to predict that the visibility will continue to decrease when $x$ increases from 10.

\subsection{The second-order interference of laser and laser light}
\begin{figure}[htb]
    \centering
    \resizebox{0.4 \textwidth}{!}
    {\includegraphics{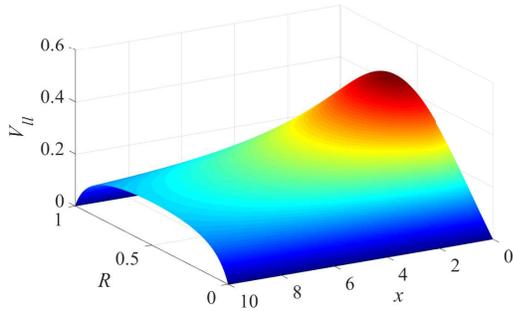}}
    \caption{Visibility of the second-order interference pattern of laser and laser light.
    }\label{ll}
\end{figure}

With the same method above, it is straightforward to calculate the second-order interference of other types of light beams. The second-order coherence function of two independent single-mode laser light beams superposed at an asymmetrical BS in Fig. \ref{setup} is
\begin{eqnarray}\label{g2-ll}
&&G^{(2)}_{ll}(\vec{r}_1,t_1;\vec{r}_2,t_2)\nonumber\\
&&=\langle|\sqrt{P_{1a}P_{2a}}e^{i\varphi_{a}}K_{a1}e^{i(\varphi_{a}'+\frac{\pi}{2})}K_{a2}\nonumber\\
&&+\sqrt{P_{1b}P_{2b}}e^{i(\varphi_{b}+\frac{\pi}{2})}K_{b1}e^{i\varphi_{b}'}K_{b2}\nonumber\\
&&+\sqrt{P_{1a}P_{2b}}e^{i\varphi_{a}''}K_{a1}e^{i\varphi_{b}''}K_{b2}\nonumber\\
&&+\sqrt{P_{1b}P_{2a}}e^{i(\varphi_{b}''+\frac{\pi}{2})}K_{b1}e^{i(\varphi_{a}''+\frac{\pi}{2})}K_{a2}|^2\rangle,
\end{eqnarray}
where the meanings of the symbols are similar as the ones in Eq. (\ref{g2-lt}). The first term on the righthand side of Eq. (\ref{g2-ll}) corresponds to two detected photons are both emitted by S$_a$. The second term on the righthand side of Eq. (\ref{g2-ll}) corresponds to two detected photons are both emitted by S$_b$. The last two terms on the righthand side of Eq. (\ref{g2-ll}) correspond to two detected photons are emitted S$_a$ and S$_b$, respectively. Since these two laser light beams are independent, Eq. (\ref{g2-ll}) can be simplified as \cite{liu-2010}
\begin{eqnarray}\label{g2-ll-1}
&&G^{(2)}_{ll}(\vec{r}_1,t_1;\vec{r}_2,t_2)\nonumber\\
&&=P_{1a}P_{2a}\langle |K_{a1}K_{a2}|^2 \rangle\nonumber\\
&&+P_{1b}P_{2b}\langle |K_{b1}K_{b2}|^2 \rangle \nonumber\\
&&+\langle |\sqrt{P_{1a}P_{2b}}K_{a1}K_{b2}-\sqrt{P_{1b}P_{2a}}K_{b1}K_{a2}|^2\rangle.
\end{eqnarray}
With the same method above \cite{liu-2013,liu-EPL,liu-submitted}, the one-dimension second-order temporal coherence function of two independent single-mode laser light beams at an asymmetrical BS in Fig. \ref{setup} is
\begin{eqnarray}\label{g2-ll-2}
&&G^{(2)}_{ll}(t_1-t_2)\nonumber\\
&&\propto 1-2\sqrt{P_{1a}P_{1b}P_{2a}P_{2b}}\cos[2\pi \Delta \nu(t_1-t_2)]
\end{eqnarray}
where all the approximations as the ones in Eq. (\ref{g2-lt-2}) have been assumed to get Eq. (\ref{g2-ll-2}). The visibility of the second-order interference pattern of two independent laser light beams is
\begin{equation}\label{vis-ll-1}
V_{ll}=\frac{2xR(1-R)}{(x+R-xR)(1-R+xR)}.
\end{equation}
Figure \ref{ll} shows how the visibility of the second-order interference pattern changes with the ratio ($x$) and the reflectivity ($R$), which is similar as the one in Fig. \ref{lt}. The visibility in these two cases is symmetrical about $R=0.5$. The visibility first increases as the $x$ increases and then decreases as $x$ continues to increase. However, there are some differences between these two situations. One difference is the maximal visibility in Fig. \ref{ll} is 0.5, while it is $1/(\sqrt{2}+1)$ in Fig. \ref{lt}. Another difference is the visibility of the second-order interference pattern of two independent single-mode laser light beams reach its maximum when $x$ equals 1 and $R$ equals 0.5, while the visibility gets it maximum when $x$ equals $\sqrt{2}/2$ and $R$ equals 0.5 for the second-order interference of laser and thermal light.

\subsection{The second-order interference of thermal and thermal light}\label{tt-subsection}
\begin{figure}[htb]
    \centering
    \resizebox{0.4 \textwidth}{!}
    {\includegraphics{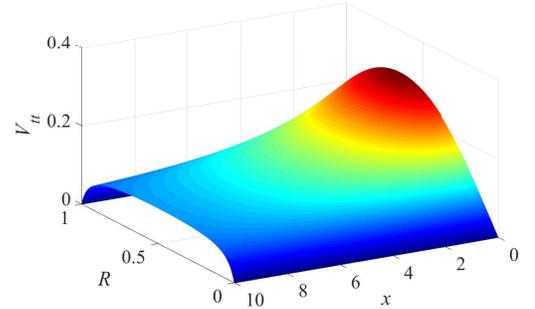}}
    \caption{Visibility of the second-order interference pattern of two independent thermal light beams.
    }\label{tt}
\end{figure}

The second-order coherence function of two independent thermal light beams at an asymmetrical BS in Fig. \ref{setup} is
\begin{eqnarray}\label{g2-tt}
&&G^{(2)}_{tt}(\vec{r}_1,t_1;\vec{r}_2,t_2)\nonumber\\
&&=\langle|\sqrt{P_{1a}P_{2a}}[\frac{1}{\sqrt{2}}e^{i\varphi_{a}}K_{a1}e^{i(\varphi_{a}'+\frac{\pi}{2})}K_{a2}\nonumber\\
&&+\frac{1}{\sqrt{2}}e^{i(\varphi_{a}+\frac{\pi}{2})}K_{a2}e^{i\varphi_{a}'}K_{a1}]\nonumber\\
&&+\sqrt{P_{1b}P_{2b}}[\frac{1}{\sqrt{2}}e^{i(\varphi_{b}+\frac{\pi}{2})}K_{b1}e^{i\varphi_{b}'}K_{b2}\nonumber\\
&&+\frac{1}{\sqrt{2}}e^{i\varphi_{b}}K_{b2}e^{i(\varphi_{b}'+\frac{\pi}{2})}K_{b1}]\nonumber\\
&&+\sqrt{P_{1a}P_{2b}}e^{i\varphi_{a}''}K_{a1}e^{i\varphi_{b}''}K_{b2}\nonumber\\
&&+\sqrt{P_{1b}P_{2a}}e^{i(\varphi_{b}''+\frac{\pi}{2})}K_{b1}e^{i(\varphi_{a}''+\frac{\pi}{2})}K_{a2}|^2\rangle,
\end{eqnarray}
where the meanings of the symbols are the same as the ones above. The visibility of the second-order interference pattern of two independent thermal light beams is
\begin{eqnarray}\label{vis-tt}
&&V_{tt}\\
&&=\frac{2xR(1-R)}{(x+R-xR)(1-R+xR)+x^2R(1-R)+R(1-R)}\nonumber.
\end{eqnarray}
Figure \ref{tt} shows how the visibility changes with the ratio ($x$) and reflectivity ($R$). The maximal visibility is $1/3$, which is gotten when $x$ equals $1$ and $R$ equals $0.5$.

\subsection{The second-order interference of photons emitted by two independent single-photon sources}\label{ss-subsection}

We have calculated the second-order interference of thermal and laser light beams at an asymmetrical BS based on two-photon interference theory. The calculation for the second-order interference of nonclassical light at an asymmetrical BS is similar as the one of classical light.  The second-order coherence function of photons emitted by two independent single-photon sources in Fig. \ref{setup} is
\begin{eqnarray}\label{g2-ss}
&&G^{(2)}_{ss}(\vec{r}_1,t_1;\vec{r}_2,t_2)\nonumber\\
&&=\langle|\sqrt{P_{1a}P_{2b}}e^{i\varphi_{a}}K_{a1}e^{i\varphi_{b}}K_{b2}\nonumber\\
&&+\sqrt{P_{1b}P_{2a}}e^{i(\varphi_{b}+\frac{\pi}{2})}K_{b1}e^{i(\varphi_{a}+\frac{\pi}{2})}K_{a2}|^2\rangle.
\end{eqnarray}
The reason why there are only two terms on the righthand side of Eq. (\ref{g2-ss}) is that single-photon source can only emit one photon at a time. The probability for two detected photons coming from the same source is zero. The only possible way to trigger a two-photon coincidence count in Fig. \ref{setup} is the detected two photons coming from S$_a$ and S$_b$, respectively. The visibility of the second-order interference pattern of photons emitted by two independent single-photon sources in Fig. \ref{setup} is
\begin{equation}\label{vis-ss}
V_{ss}=\frac{2R(1-R)}{1-2R+2R^2}.
\end{equation}

\begin{figure}[htb]
    \centering
    \resizebox{0.4 \textwidth}{!}
    {\includegraphics{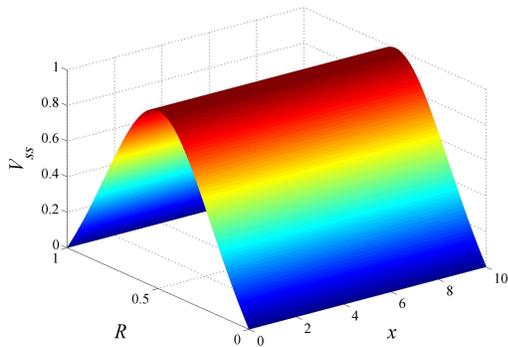}}
    \caption{Visibility of the second-order interference pattern of photons emitted by two independent single-photon sources.
    }\label{ss}
\end{figure}

Figure \ref{ss} shows the visibility of the second-order interference pattern of photons emitted by two independent single-photon sources. The maximal visibility is 1, which exceeds the limit for the visibility of the second-order interference pattern with two independent classical light beams \cite{mandel-1983}. The visibility is independent of the ratio ($x$) and only determined by the reflectivity. Since no matter what the intensity of the light emitted by single-photon source is, the two-photon coincidence counting rate is determined by the single-photon source whose emission rate is less.

\subsection{The second-order interference of photons emitted by single-photon source and single-mode laser}\label{ss-subsection}

If the photons emitted by single-photon source and single-mode laser are indistinguishable, the second-order coherence function of the photons emitted by these two sources in Fig. \ref{setup} is
\begin{eqnarray}\label{g2-sl}
&&G^{(2)}_{sl}(\vec{r}_1,t_1;\vec{r}_2,t_2)\nonumber\\
&&=\langle|\sqrt{P_{1b}P_{2b}}e^{i(\varphi_{b}+\frac{\pi}{2})}K_{b1}e^{i\varphi_{b}'}K_{b2}\nonumber\\
&&+\sqrt{P_{1a}P_{2b}}e^{i\varphi_{a}''}K_{a1}e^{i\varphi_{b}''}K_{b2}\nonumber\\
&&+\sqrt{P_{1b}P_{2a}}e^{i(\varphi_{b}''+\frac{\pi}{2})}K_{b1}e^{i(\varphi_{a}''+\frac{\pi}{2})}K_{a2}|^2\rangle,
\end{eqnarray}
where S$_a$ is assumed to be a single-photon source and S$_b$ is assumed to be a single-mode laser. The probability for the detected two photons coming from single-photon source (S$_a$) is zero. The visibility of the second-order interference pattern is
\begin{equation}\label{vis-sl}
V_{sl}=\frac{2xR(1-R)}{x(1-2R+2R^2)+R(1-R)}.
\end{equation}

\begin{figure}[htb]
    \centering
    \resizebox{0.4 \textwidth}{!}
    {\includegraphics{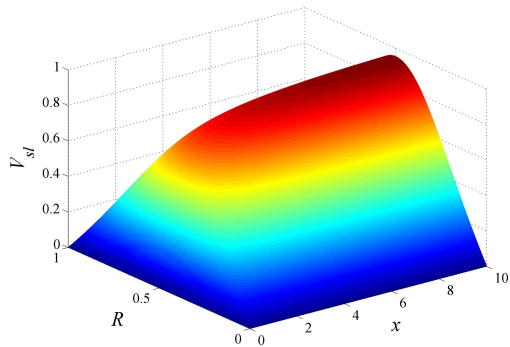}}
    \caption{Visibility of the second-order interference pattern of photons emitted single-photon source and single-mode laser.
    }\label{sl}
\end{figure}

Figure \ref{sl} shows the visibility of the second-order interference pattern of photons emitted by single-photon source and single-mode laser. Here, $x$ is the ratio between the intensities of the light emitted by single-photon source and single-mode laser. For a fixed reflectivity, visibility approaches the value in Fig. \ref{ss} as $x$ approaches infinity. Since $x$ can not be infinity in real experiment, the visibility of the second-order interference pattern of photons emitted by single-photon source and single-mode laser can not reach 1. The visibility in Fig. \ref{sl} can exceed 0.5, which is the limit for classical light \cite{mandel-1983}. The condition for $V_{sl}$ to exceed 0.5 is different for different values of $x$ and $R$. When $R$ is not in the interval of $(\frac{\sqrt{3}-1}{2\sqrt{3}},\frac{\sqrt{3}+1}{2\sqrt{3}})$, it is impossible for $V_{sl}$ exceeds 0.5 no matter what the value of the ratio might be.  When $R$ is in the interval of $(\frac{\sqrt{3}-1}{2\sqrt{3}},\frac{\sqrt{3}+1}{2\sqrt{3}})$, $x$ should be larger than $\frac{R-R^2}{6R-6R^2-1}$ in order to ensure that $V_{sl}$ exceeds 0.5.

\subsection{The second-order interference of photons emitted by single-photon source and thermal source}\label{ss-subsection}

If the photons emitted by single-photon source and thermal source are indistinguishable, the second-order coherence function of photons emitted by these two sources in Fig. \ref{setup} is
\begin{eqnarray}\label{g2-tt}
&&G^{(2)}_{st}(\vec{r}_1,t_1;\vec{r}_2,t_2)\nonumber\\
&&=\langle|\sqrt{P_{1b}P_{2b}}(\frac{1}{\sqrt{2}}e^{i(\varphi_{b}+\frac{\pi}{2})}K_{b1}e^{i\varphi_{b}'}K_{b2}\nonumber\\
&&+\frac{1}{\sqrt{2}}e^{i\varphi_{b}}K_{b2}e^{i(\varphi_{b}'+\frac{\pi}{2})}K_{b1})\nonumber\\
&&+\sqrt{P_{1a}P_{2b}}e^{i\varphi_{a}''}K_{a1}e^{i\varphi_{b}''}K_{b2}\nonumber\\
&&+\sqrt{P_{1b}P_{2a}}e^{i(\varphi_{b}''+\frac{\pi}{2})}K_{b1}e^{i(\varphi_{a}''+\frac{\pi}{2})}K_{a2}|^2\rangle,
\end{eqnarray}
where S$_a$ is assumed to be a single-photon source and S$_b$ is assumed to be a thermal source. The visibility for the second-order interference pattern is
\begin{equation}\label{vis-st}
V_{st}=\frac{2xR(1-R)}{x(1-2R+2R^2)+2R(1-R)}.
\end{equation}

\begin{figure}[htb]
    \centering
    \resizebox{0.4 \textwidth}{!}{
    \includegraphics{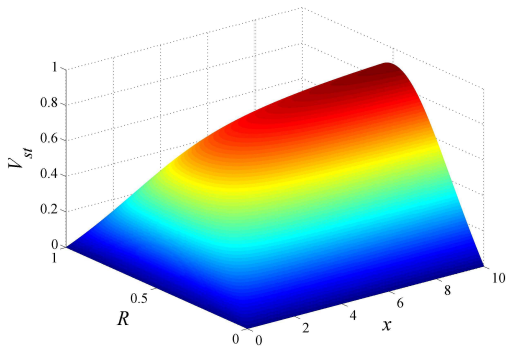}}
    \caption{Visibility of the second-order interference pattern of photons emitted single-photon source and thermal source.
    }\label{st}
\end{figure}

Figure \ref{st} shows the visibility of the second-order interference pattern, which is similar as the one in Fig. \ref{sl} except the visibility in Fig. \ref{st} is less than the one in Fig. \ref{sl} for any fixed $x$ and $R$. The visibility in Fig. \ref{st} can also exceed 0.5. When $R$ is not in the interval of $(\frac{\sqrt{3}-1}{2\sqrt{3}},\frac{\sqrt{3}+1}{2\sqrt{3}})$, it is impossible for $V_{st}$ to exceed 0.5 no matter what the value of the ratio might be.  When $R$ is in the interval of $(\frac{\sqrt{3}-1}{2\sqrt{3}},\frac{\sqrt{3}+1}{2\sqrt{3}})$, $x$ should be larger than $\frac{2R-2R^2}{6R-6R^2-1}$ in order to ensure that $V_{st}$ exceeds 0.5.

\section{Discussions}\label{discussion}

In the last section, we have calculated the visibility for the second-order interference pattern of two independent light beams at an asymmetrical BS.  Table \ref{table1} summarizes the maximal visibility of the second-order interference pattern. $l$, $t$, and $s$ are short for laser, thermal and single-photon light, respectively.  $V_{max}$ is the maximal visibility. $R_{max}$ and $x_{max}$ are the reflectivity and ratio when visibility reaches $V_{max}$, respectively.

\begin{table}[htb]
\centering
\begin{tabular}[c]{|c|c|c|c|}
\hline
$light$&$V_{max}$&$R_{max}$&$x_{max}$\\
\hline
$lt$&$1/(\sqrt{2}+1)$&$0.5$&$\sqrt{2}/2$\\
\hline
$ll$&$0.5$&$0.5$&$1$\\
\hline
$tt$&$1/3$&$0.5$&$1$\\
\hline
$ss$&$1$&$0.5$&$(0,\infty)$\\
\hline
$sl$&$\rightarrow 1$&$0.5$&$\rightarrow \infty$\\
\hline
$st$&$\rightarrow 1$&$0.5$&$ \rightarrow  \infty$\\
\hline
\end{tabular}
\caption{Maximal visibility of the second-order interference pattern of two independent light beams at an asymmetrical BS. $l$: laser light. $t$: thermal light. $s$: single-photon light. $V_{max}$: maximal visibility. $R_{max}$: reflectivity when visibility reaches $V_{max}$. $x_{max}$: ratio when visibility reaches $V_{max}$.}\label{table1}
\end{table}

It was predicted that the maximal visibility for the second-order interference pattern of two independent classical light beams is 0.5 \cite{mandel-1983}. When two independent single-mode laser light beams are superposed at a symmetrical BS, the visibility of the second-order interference pattern can reach 0.5. When two single-mode laser light beams are superposed at an asymmetrical BS ($R\neq 0.5$), the visibility is always less than 0.5. When thermal and laser light beams are superposed at BS, no matter symmetrical or asymmetrical, the visibility is less than 0.5. The maximal visibility of the second-order interference pattern of two independent thermal light beams is $1/3$. When nonclassical light is employed, the visibility of the second-order interference pattern can be larger than 0.5. For instance, the visibility can reach 1 when the photons emitted by two independent single-photon sources are superposed at a symmetrical BS. $(0,\infty)$ in row 5 of Table \ref{table1} means the visibility in this case gets its maximal for any value of $x$ when $R$ equals 0.5. The visibility of the second-order interference pattern in this case can be larger than 0.5 when the reflectivity of asymmetrical BS is in the range of ($\frac{\sqrt{3}-1}{2\sqrt{3}},\frac{\sqrt{3}+1}{2\sqrt{3}}$). The visibility can be larger than 0.5 when single-photon light is superposed with laser or thermal light. In row 6 of Table \ref{table1}, the maximal visibility approaches 1 when the ratio between the intensities of light beams emitted by single-photon source and laser approaches infinity. The visibility of the second-order interference of photons emitted by single-photon source and laser (or thermal source) is dependent on the ratio. When the visibility of the second-order interference of photons emitted by single-photon source and laser is required to be larger than 0.5, the reflectivity and ratio should satisfy some requirements as pointed out in Section 3. Only one condition is satisfied can not guarantee the visibility exceeds 0.5.

\section{Conclusions}\label{conclusion}

In conclusion, we have employed two-photon interference in Feynman's path integral theory to calculate the second-order temporal interference of two independent light beams at an asymmetrical BS. It is confirmed that the visibility of the second-order interference pattern with laser and thermal light beams can not exceed 0.5. On the other hand, when nonclassical light is employed, the visibility can be larger than 0.5. The maximal visibility of the second-order interference pattern is always got when the reflectivity of the asymmetrical BS equals 0.5, which explains why symmetrical BS instead of asymmetrical BS is employed in most of the interference experiments.

Although the conclusions in this paper are based on the second-order temporal interference, it is straightforward to generalize the method to the spatial part and prove that similar conclusions are true for the second-order spatial interference. The studies in this paper are helpful to understand the second-order interference of classical and nonclassical light at an asymmetrical BS, which is important for the applications in quantum optics and quantum information when asymmetrical BS is employed.

\section*{Acknowledgments}
This project is supported by National Science Foundation of China (No.11404255), Doctoral Fund of Ministry of Education of China (No.20130201120013), the 111 Project of China (No.B14040) and the Fundamental Research Funds for the Central Universities.

\end{document}